\RequirePackage{fix-cm}
\documentclass[smallextended]{svjour3}       % onecolumn (second format)
\smartqed  % flush right qed marks, e.g. at end of proof
\usepackage{graphicx}
 \usepackage{mathptmx}      % use Times fonts if available on your TeX system
\journalname{Hyperfine Interactions}
\begin{document}

\title{Muon decay in orbit spectra for $\mu-e$ conversion experiments%\thanks{Grants or other notes
}


\author{Andrzej Czarnecki         \and
        Xavier Garcia i Tormo \and William J. Marciano%etc.
}


\institute{A. Czarnecki \and X. Garcia i Tormo \at
              Department of Physics, University of Alberta, Edmonton, Alberta,
Canada T6G 2E1
                     %  \\
           \and
          X. Garcia i Tormo \at
           Institut f\"ur Theoretische Physik, Universit\"at Bern,
  Sidlerstrasse 5, CH-3012 Bern, Switzerland
            \and
           William J. Marciano \at
           Department of Physics, Brookhaven National Laboratory, Upton, NY
11973, USA
}

\date{Received: date / Accepted: date}

\maketitle

\begin{abstract}
We have determined in detail the electron spectrum in the decay of bound
muons. These results are especially relevant for the upcoming $\mu-e$
conversion experiments.
%\keywords{Bound muon decay \and $\mu-e$ conversion}
\end{abstract}

\section{Introduction}
From the observation of neutrino oscillations, we now know that lepton
flavors are not conserved. However,
the mixing and small neutrino mass differences seen in oscillations
have a negligible effect on charged-lepton flavor violating (CLFV) reactions. 
Thus, the CLFV
reactions provide a discovery window for interactions beyond Standard
Model expectations \cite{Kuno:1999jp,LeptMomBook}.

Muons play a central role in searches for CLFV 
\cite{Kuno:1999jp,LeptMomBook}, because they can be produced in large numbers 
and live relatively long.
One reaction that can be probed with particularly high sensitivity
is coherent muon-electron conversion in a muonic atom, 
\begin{equation}
\mu^{-}+(A,Z)\to e^{-}+(A,Z),\label{eq:mueconv}
\end{equation}
where $(A,Z)$ represents a nucleus of atomic number $Z$ and mass
number $A$. It has the advantage of producing just a single particle, a mono-energetic electron.  It  does not have the problem of accidental background  that plagues searches for the decay $\mu^+\to e^+\gamma$, which can be mimicked  by a positron from a normal muon decay and a photon
coming from the radiative decay of a different muon, bremsstrahlung,
or positron annihilation-in-flight.
Various experiments have been performed over the years
to search for the  conversion \cite{Marciano:2008zz}. The most stringent results 
come from the SINDRUM II Collaboration \cite{Bertl:2006up},
which reports an upper limit of $7\times10^{-13}$ for the branching
ratio of the conversion process relative to muon capture in gold. Several
new efforts are being planned. In the nearest future, the DeeMe
Collaboration \cite{DeeMe} has proposed to reach $10^{-14}$ sensitivity. Larger scale searches,
Mu2e at Fermilab \cite{Carey:2008zz} and COMET at J-PARC \cite{Cui:2009zz},
aim for sensitivities below $10^{-16}$. In the long run, intensity
upgrades at Fermilab and the proposal PRISM/PRIME at
J-PARC may allow them to reach $10^{-18}$ sensitivity. A quite remarkable improvement
of about four orders of magnitude, with respect to the current limit, is
therefore envisaged.

The success of the conversion searches depends critically on  control
of the background events. The signal for the $\mu-e$ conversion process
in Eq.~(\ref{eq:mueconv}) is a mono-energetic electron with energy
$E_{\mu e}$, given by 
\begin{equation}
E_{\mu e}=m_{\mu}-E_{\mathrm{b}}-E_{\mathrm{rec}},\label{eq:conven}
\end{equation}
where $m_{\mu}$ is the muon mass, $E_{\mathrm{b}}\simeq Z^2\alpha^2m_{\mu}/2$ is the binding
energy of the muonic atom, and $E_{\mathrm{rec}}\simeq m_{\mu}^2/(2m_N)$ is the nuclear-recoil
energy, with $\alpha$ the fine-structure constant and $m_N$ the
nucleus mass. The main physics background for this signal comes from the so-called
muon decay in orbit (DIO), a process in which the muon decays in
the normal way,  $\mu^{-}\to e^{-}\overline{\nu}_{e}\nu_{\mu}$, while
in the orbit of the atom. Whereas in a free muon decay, in order to
conserve energy and three-momentum, the maximum electron energy is
$m_{\mu}/2$, for DIO, the nucleus recoil can balance the electron's
three-momentum taking basically no energy. This allows for the
maximum electron energy to be $E_{\mu e}$, close to the full muon mass $m_\mu$.  
Therefore, the high-energy  electrons from the muon decay in orbit constitute a
background for conversion searches.

\section{Muon decay in orbit}
Several theoretical studies of the muon decay in orbit have been
published. Expressions describing the electron spectrum including relativistic effects in the muon
wavefunction, the Coulomb interaction between the electron and the
nucleus and a finite nuclear size have been available for some time
\cite{Haenggi:1974hp,Watanabe:1987su,Watanabe:1993}. However, the high-energy
endpoint of the spectrum, which is the most important region for conversion-search experiments, was not studied in detail. 
References \cite{Shanker:1981mi,Shanker:1996rz} did study the high-energy
end of the electron spectrum, and presented approximate results which
allow for a quick rough estimate of the muon decay in orbit contribution
to the background in conversion experiments. We have performed a new
evaluation of the DIO spectrum, considering in detail all the 
effects  needed in the high-energy region \cite{Czarnecki:2011mx}.
Our results describe the background contribution for $\mu-e$
conversion searches, as well as a check on previous low- and
high-energy partial calculations \cite{Watanabe:1993,Shanker:1981mi}
and an interpolation between them. It is worth emphasizing that not
only the high-energy region is relevant for conversion experiments, but
the full spectrum is necessary in order to study reconstruction errors
in the detector. 

To obtain the correct result for the high-energy tail of the spectrum
it is crucial to include nuclear-recoil effects, since they modify the
endpoint energy (see Eq.~(\ref{eq:conven})). Also, to produce an
on-shell electron with energy around $m_{\mu}$, either the muon must be at the 
tail of the bound-state wavefunction or the produced electron must interact
with the nucleus. This tells us that the full Dirac
equation for the muon as well as the interaction of
the outgoing electron with the field of the nucleus must be taken into
account. Also, finite-nuclear-size effects will be most
important in this region. Order $\alpha$ radiative
corrections are not expected to significantly modify the results at
the endpoint, and are not included in our results. Uncertainties in
the modelling of finite nuclear-size effects induce errors in the
spectrum that increase as we approach the endpoint, but those errors are never
larger than a few percent. 

\section{Results and discussion}
\begin{figure}
 \includegraphics[width=\textwidth]{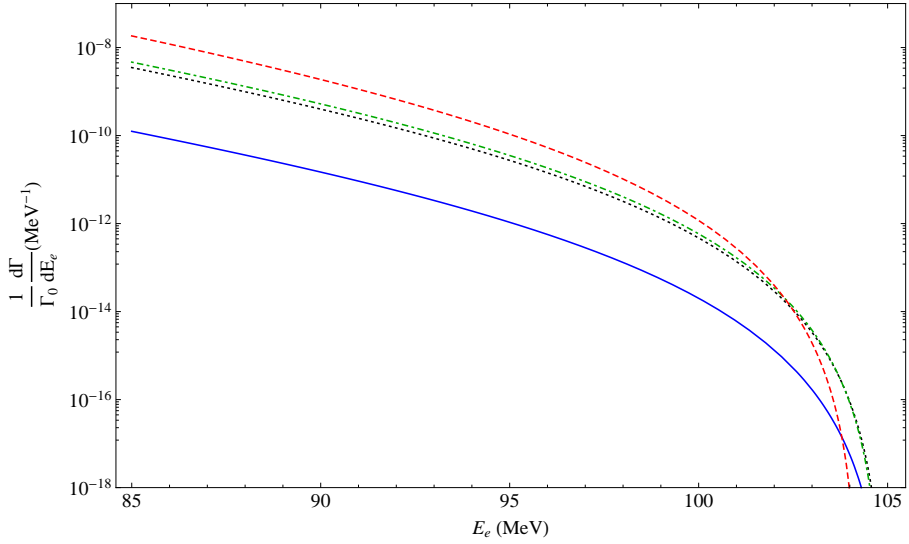}
\caption{Electron spectrum, normalized to the free-muon decay rate
  $\Gamma_0$. The solid blue line is for carbon, the black dotted line
for aluminum, the green dot-dashed line for silicon and the red dashed
line for titanium.}
\label{fig:eplog}       
\end{figure}

\begin{figure}
 \includegraphics[width=0.49\textwidth]{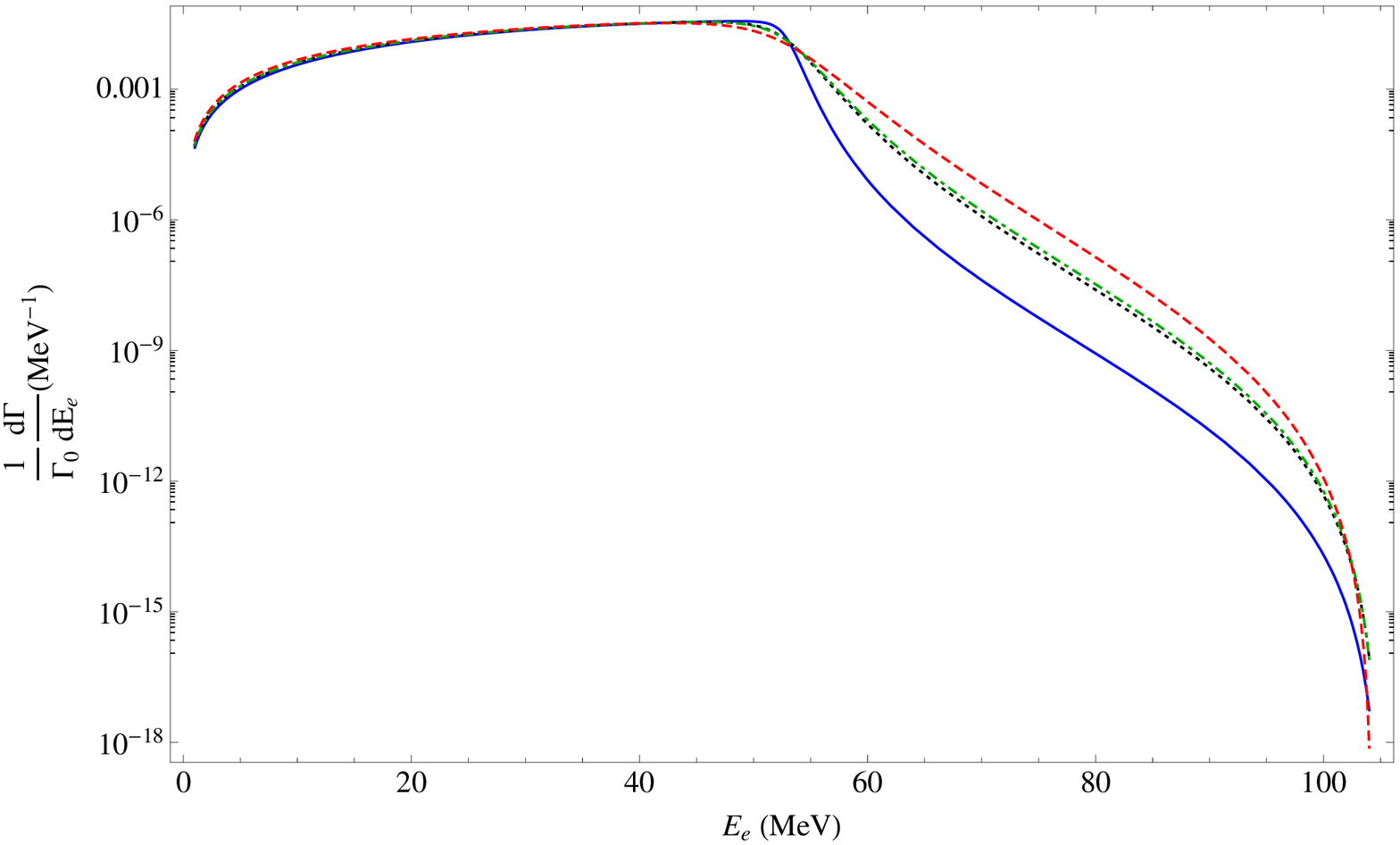}
  \includegraphics[width=0.49\textwidth]{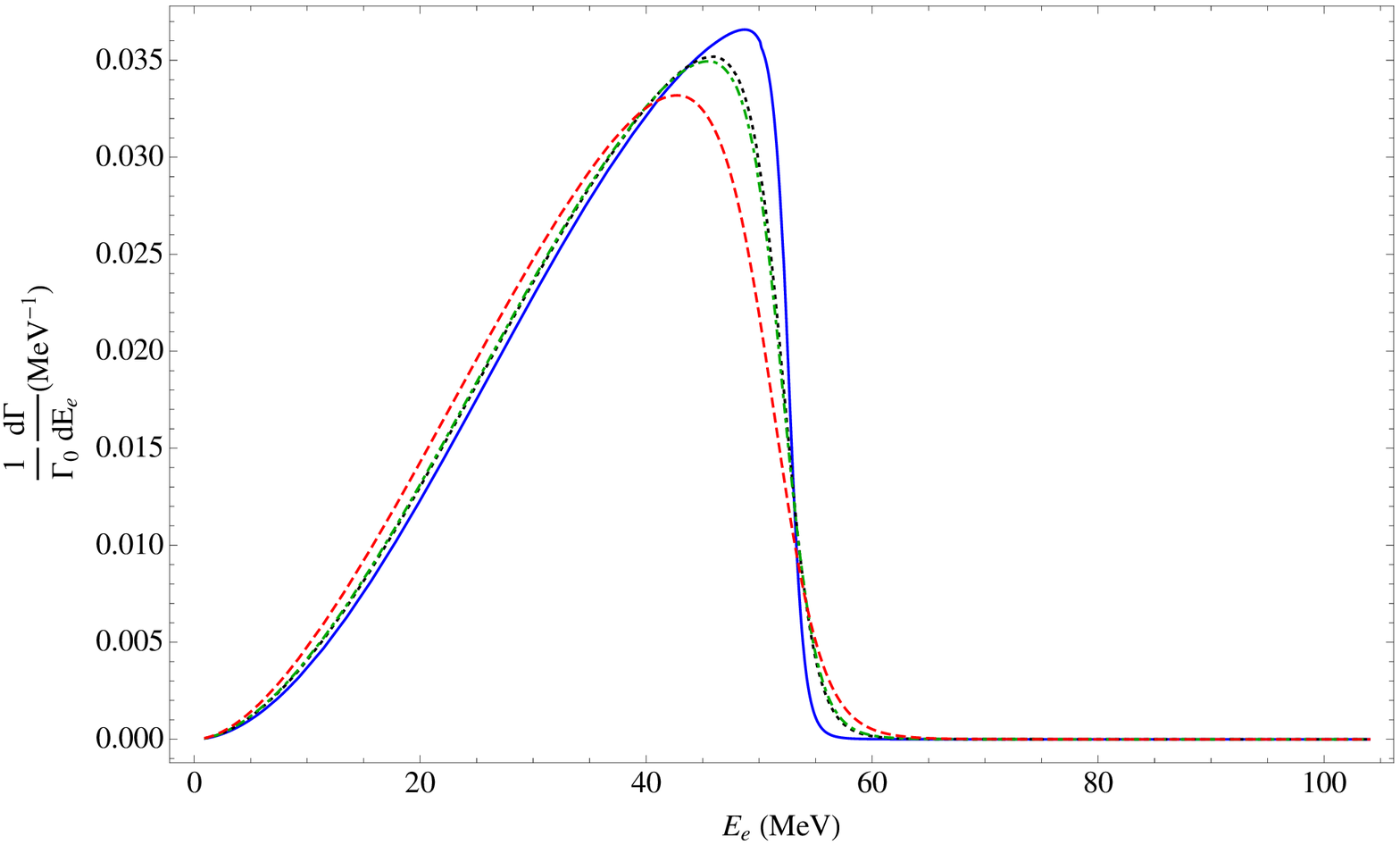}
\caption{Electron spectrum for the full range of $E_e$; see caption of Fig.~\ref{fig:eplog}.}
\label{fig:full}       
\end{figure}

Here we present our results for the elements that are relevant for the
upcoming conversion experiments. The DeeMe Collaboration plans to use a
silicon-carbide target, whereas the Mu2e and COMET Collaborations are
considering aluminum and titanium as targets. In Table  \ref{tab:en} we give the values of
the bound muon energy $E_{\mu}=m_{\mu}-E_{\mathrm{b}}$ and the electron
endpoint energy $E_{\mu e}$, for carbon, aluminum, silicon and titanium.
\begin{table}
\caption{Values for muon energies $E_{\mu}$, nuclear masses $m_N$, and
  endpoint energies $E_{\mu e}$.}
\label{tab:en}
\begin{tabular}{cllll}
\hline\noalign{\smallskip}
Nucleus & Z & $E_{\mu}$ (MeV) & $m_N$ (MeV )& $E_{\mu e}$ (MeV)  \\
\noalign{\smallskip}\hline\noalign{\smallskip}
C & 6 & 105.557 & 11188 & 105.06 \\
Al &13 & 105.194 & 25133 & 104.973 \\
Si & 14 & 105.121 & 26162 & 104.91 \\
Ti & 22 & 104.394 & 44588 & 104.272 \\
\noalign{\smallskip}\hline
\end{tabular}
\end{table}
We present the results of the numerical evaluation of the spectra for
those elements in Figs.~\ref{fig:eplog} and \ref{fig:full}.
Regarding the nuclear distributions, a two-parameter Fermi
distribution has been used for aluminum and titanium, a
three-parameter Fermi distribution for silicon and a Fourier-Bessel
expansion for carbon \cite{Czarnecki:2011mx,De Jager:1987qc,Fricke:1995zz}.
Those results are useful for assessing the DIO background events from carbon for the
DeeMe experiment which will search primarily for conversion in silicon.  
They also illustrate the electron resolution requirements, as a function of stopping target,
needed to reach future high sensitivity goals.

The high sensitivity that the upcoming conversion experiments will
reach may also allow them to improve the present bounds on some exotic muon
decays, like the decay of the muon into an electron and a majoron (a
Goldstone boson that appears in models where lepton number is a
spontaneously broken global symmetry) \cite{Tormo:2011et}.

\begin{acknowledgements}
This research was supported by Science and
Engineering Research Canada (NSERC) and by the United States Department of
Energy under Grant Contract DE-AC02-98CH10886.
\end{acknowledgements}



\end{document}